\documentclass[12pt,fleqn]{article}
\usepackage{palatino}
\usepackage{graphicx}
\usepackage[round]{natbib}
\def\alt{\mathrel{\hbox{\rlap{\hbox{\lower4pt\hbox{$\sim$}}}
\hbox{\raisebox{0.4ex}{\hspace*{-0.05in}$<$}}}}}

\def\agt{\mathrel{\hbox{\rlap{\hbox{\lower4pt\hbox{$\sim$}}}
\hbox{\raisebox{0.4ex}{\hspace*{-0.05in}$>$}}}}}
\begin{document}
 
\leftline{\Large\bf Analyzing neural responses to natural signals:}
\leftline{\Large\bf Maximally informative dimensions}
\bigskip\bigskip

\leftline{\large Tatyana Sharpee,$^1$ Nicole C. Rust,$^2$ and William
Bialek$^{1,3}$}
\bigskip

\noindent $^1$ Sloan--Swartz Center for Theoretical Neurobiology and Department of\\
Physiology, University of California at San Francisco, San Francisco,
CA 94143\\
$^2$ Center for Neural Science, New York University, New York, NY 10003\\
$^3$ Department of Physics, Princeton University, Princeton, New Jersey 08544\\
{\it sharpee@phy.ucsf.edu, rust@cns.nyu.edu, wbialek@princeton.edu}
\bigskip\bigskip

\leftline{\today} 

\bigskip\bigskip\hrule\bigskip\bigskip

\begin{quote}  
  We propose a method that allows for a rigorous statistical analysis
  of neural responses to natural stimuli which are non--Gaussian and
  exhibit strong correlations.  We have in mind a model in which
  neurons are selective for a small number of stimulus dimensions out
  of a high dimensional stimulus space, but within this subspace the
  responses can be arbitrarily nonlinear. Existing analysis methods
  are based on correlation functions between stimuli and responses,
  but these methods are guaranteed to work only in the case of
  Gaussian stimulus ensembles. As an alternative to correlation
  functions, we maximize the mutual information between the neural
  responses and projections of the stimulus onto low dimensional
  subspaces. The procedure can be done iteratively by increasing the
  dimensionality of this subspace. Those dimensions that allow the
  recovery of all of the information between spikes and the full
  unprojected stimuli describe the relevant subspace. If the
  dimensionality of the relevant subspace indeed is small, it becomes
  feasible to map the neuron's input--output function even under fully
  natural stimulus conditions. These ideas are illustrated in
  simulations on model visual and auditory neurons responding to
  natural scenes and sounds, respectively.
\end{quote}

\section{Introduction}
\label{sec:intro}
From olfaction to vision and audition, a growing number of experiments
are examining the responses of sensory neurons to natural stimuli
\citep{Creutzfeldt78,Rieke95,Baddeley97,Stanley99,Theunissen00,Vinje00,Lewen01,Sen01,Vickers01,Vinje02,Ringach02,Weliky03,Rolls03,Smyth03}.
Observing the full dynamic range of neural responses may require using
stimulus ensembles which approximate those
 occurring in nature \citep{Rieke_book,Simoncelli01}, and it is an
attractive hypothesis that the neural representation of these natural
signals may be optimized in some way
\citep{Barlow61,Barlow01,Twer01,leshouches_notes}.  Many neurons
exhibit strongly nonlinear and adaptive responses that are unlikely to
be predicted from a combination of responses to simple stimuli; for
example neurons have been shown to adapt to the distribution of
sensory inputs, so that any characterization of these responses will
depend on context \citep{Smirnakis96,brenner00-adapt,Fairhall01}.  Finally, the
variability of a neuron's responses decreases substantially when complex
dynamical, rather than static, stimuli are used
\citep{MainenSejnowski,deRuyter97,Kara00,canberra}. All of these
arguments point to the need for general tools to analyze the neural
responses to complex, naturalistic inputs.

\begin{figure}[t]
\begin{center}
\includegraphics[width=4.0in]{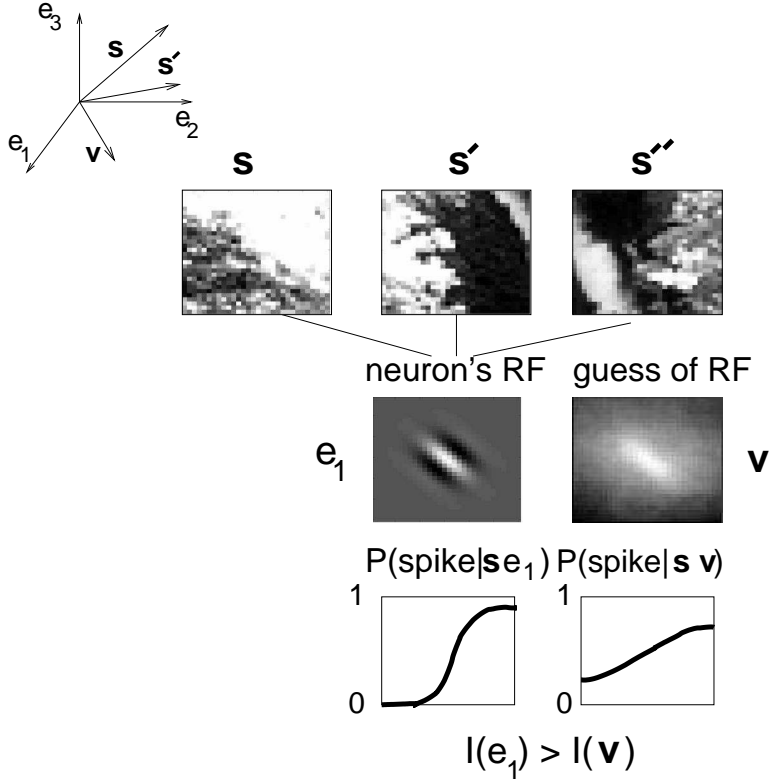}
\end{center}
\vspace*{-0.2in}
\caption{Schematic illustration of a model with a one--dimensional
relevant subspace: $\hat e_1$ is the relevant dimension, and $\hat
e_2$ and $\hat e_3$ are irrelevant ones.  Shown below are three example
stimuli, ${\bf s}$, ${\bf s}'$, and ${\bf s}''$, the receptive field of a model
neuron---the relevant dimension $\hat e_1$, and our guess ${\bf v}$ for
the relevant dimension. Probabilities of a spike $P({\rm spike}|{\bf s}
\cdot \hat e_1)$ and $P({\rm spike}| {\bf s }\cdot {\bf v})$ are calculated
by first projecting all of the stimuli ${\bf s}$ onto each of the two
vectors $\hat e_1$ and ${\bf v}$, respectively, and then applying
Eqs.~(\protect\ref{histo},\ref{bayes},\ref{ior}) sequentially. Our
guess ${\bf v}$ for the relevant dimension is adjusted during the
progress of the algorithm in such a way as to maximize $I({\bf v})$ of
Eq.~(\protect\ref{Iv}), which makes vector ${\bf v}$ approach the true
relevant dimension ${\hat e_1}$. }
\label{fig:illustr}
\vspace*{-0.1in}
\end{figure}

The stimuli analyzed by sensory neurons are intrinsically high
dimensional, with dimensions $D \sim 10^2 - 10^3$. For example, in the
case of visual neurons, the input is commonly specified as light
intensity on a grid of at least $10\times 10$ pixels. Each of the
presented stimuli can be described as a vector ${\bf s}$ in this high
dimensional stimulus space, see Fig.~\ref{fig:illustr}. The dimensionality becomes even larger if
stimulus history has to be considered as well. For example, if we are
interested in how the past $N$ frames of the movie affect the
probability of a spike, then the stimulus ${\bf s}$, being a concatenation
of the past $N$ samples, will have dimensionality $N$ times that of a
single frame.  We also assume that the probability distribution
$P({\bf s})$ is sampled during an experiment ergodically, so that we
can exchange averages over time with averages over the true
distribution as needed.

Even though direct exploration of a $D \sim 10^2 - 10^3$ dimensional
stimulus space is beyond the constraints of experimental data
collection, progress can be made provided we make certain assumptions
about how the response has been generated.  In the simplest model, the
probability of response can be described by one receptive field (RF)
or linear filter \citep{Rieke_book}.  The receptive field can be
thought of as a template or special direction $\hat e_1$ in the
stimulus space\footnote{\normalsize The notation $\hat e$ denotes a unit vector,
since we are interested only in the direction the vector specifies and
not in its length.} such that the neuron's response depends only on a
projection of a given stimulus ${\bf s}$ onto $\hat e_1$, although the
dependence of the response on this projection can be strongly
nonlinear, cf. Fig.~\ref{fig:illustr}. In this simple model, the
reverse correlation method \citep{deBoer,Rieke_book,Chichilnisky01}
can be used to recover the vector $\hat e_1 $ by analyzing the
neuron's responses to Gaussian white noise.  In a more general case,
the probability of the response depends on projections $s_i= \hat e_i
\cdot {\bf s}$ of the stimulus ${\bf s}$ on a set of $K$ vectors $\{
\hat e_1, \,\hat e_2,\, ...\, ,\hat e_K \}$:
\begin{equation}
\label{ior}
P({\rm spike}|{\bf s})=P({\rm spike}) f(s_1,s_2, ...,s_K), 
\end{equation}
where $P({\rm spike}|{\bf s})$ is the probability of a spike given a
stimulus ${\bf s}$ and $P({\rm spike})$ is the average firing rate. In
what follows we will call the subspace spanned by the set of vectors
$\{ \hat e_1, \,\hat e_2,\, ...\, ,\hat e_K \}$ the relevant subspace
(RS)\footnote{\normalsize Since the analysis does not depend on a
particular choice of a basis within the full D-dimensional stimulus
space, for clarity we choose the basis in which the first $K$ basis
vectors span the relevant subspace and the remaining $D-K$ vectors
span the irrelevant subspace.}. We reiterate that vectors $\{\hat e_i
\}$, $1\leq i \leq K$ may also describe how the time dependence of the
stimulus ${\bf s}$ affects the probability of a spike. An example of
such a relevant dimension would be a spatiotemporal receptive field of
a visual neuron.  Even though the ideas developed below can be used to
analyze input--output functions $f$ with respect to different neural
responses, such as patterns of spikes in time
\citep{stcov,B00a,Reinagel00}, for illustration purposes we choose a
single spike as the response of interest.\footnote{ {\normalsize We
emphasize that our focus here on single spikes is {\em not}}
{\normalsize equivalent to assuming that the spike train is a Poisson
process modulated by the stimulus.  No matter what the statistical
structure of the spike train is we always can ask what features of the
stimulus are relevant for setting the probability of generating a
single spike at one moment in time.  From an information theoretic
point of view, asking for stimulus features that capture the mutual
information between the stimulus and the arrival times of single
spikes is a well posed question even if successive spikes do not carry
independent information; note also that spikes carrying independent
information is not the same as spikes being generated as a Poisson
process.  On the other hand, if (for example) different temporal
patterns of spikes carry information about different stimulus
features, then analysis of single spikes will result in a relevant
subspace of artefactually high dimensionality.  Thus it is important
that the approach discussed here carries over without modification to
the analysis of relevant dimensions for the generation of any discrete
event, such as a pattern of spikes across time in one cell or
synchronous spikes across a population of cells.  For a related
discussion of relevant dimensions and spike patterns using covariance
matrix methods see \citep{stcov,Arcas03}.  }}

Equation (\ref{ior}) in itself is not yet a simplification if the
dimensionality $K$ of the RS is equal to the dimensionality $D$ of the
stimulus space.  In this paper we will assume that the neuron's firing
is sensitive only to a small number of stimulus features, i.e. $K \ll
D$.  While the general idea of searching for low dimensional structure
in high dimensional data is very old, our motivation here comes from
work on the fly visual system where it was shown explicitly that
patterns of action potentials in identified motion sensitive neurons
are correlated with low dimensional projections of the high
dimensional visual input
\citep{stcov,brenner00-adapt,rob+bill-feature}. The input--output
function $f$ in Eq.~(\ref{ior}) can be strongly nonlinear, but it is
presumed to depend only on a small number of projections. This
assumption appears to be less stringent than that of approximate
linearity which one makes when characterizing neuron's response in
terms of Wiener kernels [see for example the discussion in Section
2.1.3 of \cite{Rieke_book}].  The most difficult part in
reconstructing the input--output function is to find the RS. Note that
for $K>1$, a description in terms of any linear combination of vectors
$\{ \hat e_1, \,\hat e_2,\, ...\, ,\hat e_K \}$ is just as valid,
since we did not make any assumptions as to a particular form of the
nonlinear function $f$.

Once the relevant subspace is known, the probability $P({\rm
  spike}|{\bf s})$ becomes a function of only a few parameters, and it
becomes feasible to map this function experimentally, inverting the
probability distributions according to Bayes' rule:
\begin{equation}
\label{bayes}
 f(s_1,s_2, ...,s_K)=\frac{P(s_1,s_2, ...,s_K|{\rm spike})}{P(s_1,s_2, ...,s_K )}.
\end{equation}
If stimuli are chosen from a correlated Gaussian noise ensemble, then
the neural response can be characterized by the spike--triggered
covariance method
\citep{stcov,brenner00-adapt,Schwartz02,Touryan02,rob+bill-feature}. It can be
shown that the dimensionality of the RS is equal to the number of
nonzero eigenvalues of a matrix given by a difference between
covariance matrices of all presented stimuli and stimuli conditional
on a spike. Moreover, the RS is spanned by the eigenvectors associated
with the nonzero eigenvalues multiplied by the inverse of the {\it a
priori} covariance matrix.  Compared to the reverse correlation
method, we are no longer limited to finding only one of the relevant
dimensions $\{\hat e_i \}$, $1 \leq i \leq K$. Both the reverse
correlation and the spike--triggered covariance method, however, give
rigorously interpretable results {\it only} for Gaussian distributions
of inputs.

In this paper we investigate whether it is possible to lift the
requirement for stimuli to be Gaussian. When using natural stimuli,
which certainly are non--Gaussian, the RS cannot be found by the
spike--triggered covariance method.  Similarly, the reverse
correlation method does not give the correct RF, even in the simplest
case where the input--output function in Eq. (\ref{ior}) depends only
on one projection; see Appendix~\ref{app:non} for a discussion of this
point.  However, vectors that span the RS are clearly special
directions in the stimulus space independent of assumptions about
$P({\bf s})$.  This notion can be quantified by the Shannon
information.  We note that the stimuli ${\bf s}$ do not have to lie on
a low-dimensional manifold within the overall $D$ dimensional
space.\footnote{ \normalsize If one suspects that neurons are sensitive to low
dimensional features of their input, one might be tempted to analyze
neural responses to stimuli that explore only the (putative) relevant
subspace, perhaps along the line of the subspace reverse correlation
method \citep{Ringach97}.  Our approach (like the spike--triggered
covariance approach) is different because it allows the analysis of
responses to stimuli that live in the full space, and instead we let
the neuron ``tell us'' which low dimensional subspace is relevant.}
However, since we assume that the neuron's input-output function
depends on a small number of relevant dimensions, the ensemble of
stimuli conditional on a spike may exhibit clear clustering. This
makes the proposed method of looking for the RS complimentary to the
clustering of stimuli conditional on a spike done in the information
bottleneck method \citep{Tishby}; see also
\citep{JPMiller01}. Non--information based measures of similarity
between probability distributions $P({\bf s})$ and $P({\bf s}|{\rm
spike})$ have also been proposed to find the RS
\citep{Paninski_NIPS03}.

\noindent
To summarize our assumptions:
\begin{itemize}
\item{The sampling of the probability distribution of stimuli $P({\bf
s})$ is ergodic and stationary across repetitions. The probability
distribution is not assumed to be Gaussian. The ensemble of stimuli
described by $P({\bf s})$ does not have to lie on a low-dimensional
manifold embedded in the overall $D$-dimensional space.}
\item{We choose a single spike as the response of interest (for
illustration purposes only). An identical scheme can be applied, for
example, to particular interspike intervals or to synchronous spikes
from a pair of neurons.}
\item{The subspace relevant for generating a spike is low dimensional
and Euclidean, cf. Eq.~\ref{ior}.}
 \item{The input--output function, which is defined within the low
 dimensional RS, can be arbitrarily nonlinear. It is obtained
 experimentally by sampling the probability distributions $P({\bf s})$
 and $P({\bf s} |{\rm spike})$ within the RS.}
\end{itemize}

The paper is organized as follows.  In Sec.~\ref{sec:info} we discuss
how an optimization problem can be formulated to find the RS.  A
particular algorithm used to implement the optimization scheme is
described in Sec.~\ref{sec:algorithm}. In Sec.~\ref{sec:results} we
illustrate how the optimization scheme works with natural stimuli for
model orientation sensitive cells with one and two relevant
dimensions, much like simple and complex cells found in primary visual
cortex, as well as for a model auditory neuron responding to natural
sounds. We also discuss the convergence of our estimates of the RS as
a function of data set size.  We emphasize that our optimization
scheme does not rely on any specific statistical properties of the
stimulus ensemble, and thus can be used with natural stimuli.

\section{Information as an objective function} 
\label{sec:info}

When analyzing neural responses, we compare the {\it a priori}
probability distribution of all presented stimuli with the probability
distribution of stimuli which lead to a spike \citep{stcov}.  For
Gaussian signals, the probability distribution can be characterized by
its second moment, the covariance matrix. However, an ensemble of
natural stimuli is not Gaussian, so that in a general case neither
second nor any finite number of moments is sufficient to describe the
probability distribution.  In this situation, Shannon information
provides a rigorous way of comparing two probability
distributions. The average information carried by the arrival time of
one spike is given by \citep{B00a}
\begin{equation}
\label{ispike}
I_{\rm spike}= \int d{\bf s} P({\bf s}|{\rm spike}) \log_2 \left[
{{P({\bf s}|{\rm spike})}\over
{P({\bf s})}}\right]\,,
\end{equation}
where $d{\bf s}$ denotes integration over full $D$--dimensional stimulus
space. The information per spike as written in (\ref{ispike}) is
difficult to estimate experimentally, since it requires either
sampling of the high--dimensional probability distribution $P({\bf
s}|{\rm spike})$ or a model of how spikes were generated, i.e. the
knowledge of low--dimensional RS. However it is possible to calculate
$I_{\rm spike}$ in a model--independent way, if stimuli are presented
multiple times to estimate the probability distribution $ P({\rm
spike}|{\bf s})$. Then,
\begin{equation}
\label{exp_ispike}
I_{\rm spike}=\left\langle \frac{P({\rm spike}|{\bf s})}{P({\rm spike})} \log_2\left[
\frac{P({\rm spike}|{\bf s})}{P({\rm spike})}\right]\right\rangle_{\bf s}, 
\end{equation}
where the average is taken over all presented stimuli. This can be
useful in practice \citep{B00a}, because we can replace the ensemble
average $\langle \rangle_{\bf s}$ with a time average, and $P({\rm
spike}|{\bf s})$ with the time dependent spike rate $r(t)$.  Note that
for a finite dataset of $N$ repetitions, the obtained value $I_{\rm
spike}(N)$ will be on average larger than $ I_{\rm
spike}(\infty)$. The true value $I_{\rm spike}$ can be found either by
subtracting an expected bias value, which is of the order of $\sim
1/(P({\rm spike})N \, 2 \ln 2)$
\citep{Treves,Panzeri96,Pola02,Paninski_NC03}, or by extrapolating to
$N\to\infty$ \citep{B00a,entropy}. Measurement of $I_{\rm spike}$ in
this way provides a model independent benchmark against which we can
compare any description of the neuron's input--output relation.

Our assumption is that spikes are generated according to a projection
onto a low dimensional subspace. Therefore to characterize relevance
of a particular direction ${\bf v}$ in the stimulus space, we project
all of the presented stimuli onto ${\bf v}$ and form probability
distributions $ P_{\bf v}(x)$ and $ P_{\bf v}(x|{\rm spike})$ of
projection values $x$ for the {\it a priori} stimulus ensemble and
that conditional on a spike, respectively:
\begin{eqnarray}
\label{histo}
  P_{\bf v}(x)&=&\langle\delta(x-{\bf s}\cdot {\bf v})\rangle_{\bf s},\\
 P_{\bf v}(x|{\rm spike})&=&\langle\delta(x-{\bf s}\cdot{\bf v})|{\rm spike}\rangle_{\bf
  s},
\end{eqnarray}
where $\delta(x)$ is a delta-function. In practice, both the average
  $\langle \cdots \rangle_{\bf s}\equiv \int d{\bf s} \cdots P({\bf
  s})$ over the {\it a priori} stimulus ensemble, and the average
  $\langle \cdots | {\rm spike}\rangle_{\bf s} \equiv\int d{\bf s}
  \cdots P({\bf s}|\rm spike)$ over the ensemble conditional on a
  spike, are calculated by binning the range of projections values
  $x$. The probability distributions are then obtained as histograms,
  normalized in a such a way that the sum over all bins gives 1.  The mutual information between spike arrival times and the projection $x$, by analogy with Eq. (\ref{ispike}), is 
\begin{equation}
\label{Iv}
I( {\bf v})=\int dx P_{\bf v}(x|{\rm spike})\log_2\left[{{P_{\bf v}(x|{\rm spike})}\over
{P_{\bf v}(x)}}\right],
\end{equation}
which is also the Kullback-Leibler divergence $D\left[P_{\bf
v}(x|{\rm spike}) || P_{\bf v}(x)\right]$; notice that this information is a function of the direction ${\bf v}$.  The information $I({\bf v})$
provides an invariant measure of how much the occurrence of a spike is
determined by projection on the direction ${\bf v}$.  It is a function
only of direction in the stimulus space and does not change when
vector ${\bf v}$ is multiplied by a constant.  This can be seen by
noting that for any probability distribution and any constant $c$,
$P_{c{\bf v}}(x)=c^{-1}P_{\bf v}(x/c)$ [see also Theorem 9.6.4 of
\cite{Cover_Thomas}].  When evaluated along any vector ${\bf v}$, $I({\bf
v})\leq I_{\rm spike}$.  The total information $I_{\rm spike}$ can be
recovered along one particular direction only if ${\bf v}=\hat e_1$,
and only if the RS is one dimensional.

By analogy with (\ref{Iv}), one could also calculate information
$I({\bf v}_1,... ,{\bf v}_n)$ along a set of several directions
$\{{\bf v}_1,... , {\bf v}_n \}$ based on the multi-point probability
distributions of projection values $x_1$, $x_2$, ... $x_n$ along
vectors ${\bf v}_1$, ${\bf v}_2$, ... ${\bf v}_n$ of interest:
\begin{eqnarray}
P_{{\bf v}_1,...,{\bf v}_n}(\{x_i\}|{\rm spike})&=&
\Bigg\langle \prod_{i=1}^n\delta(x_i-{\bf s}\cdot {\bf v}_i) | {\rm spike} \Bigg\rangle_{\bf s}, \\
 P_{{\bf v}_1,...,{\bf v}_n}(\{x_i\})&=&\Bigg\langle\prod_{i=1}^n\delta(x_i-{\bf s}\cdot {\bf v}_i)\Bigg\rangle_{\bf s}.
\end{eqnarray}

If we are successful in finding all of the directions $\{ \hat e_i
\}$, $1\leq i\leq K$ contributing to the input--output relation
(\ref{ior}), then the information evaluated in this subspace will be
equal to the total information $I_{\rm spike}$. When we calculate
information along a set of $K$ vectors that are slightly off from the
RS, the answer, of course, is smaller than $I_{\rm spike}$ and is
initially quadratic in small deviations $\delta {\bf v}_i$.  One can
therefore hope to find the RS by maximizing information with respect
to $K$ vectors simultaneously.  The information does not increase if
more vectors outside the RS are included.  For uncorrelated stimuli,
any vector or a set of vectors that maximizes $I({\bf v})$ belongs to
the RS.  On the other hand, as discussed in Appendix~\ref{app:max},
the result of optimization with respect to a number of vectors $k<K$
may deviate from the RS if stimuli are correlated.  To find the RS, we
first maximize $I({\bf v})$, and compare this maximum with $I_{\rm
spike}$, which is estimated according to (\ref{exp_ispike}). If the
difference exceeds that expected from finite sampling corrections, we
increment the number of directions with respect to which information
is simultaneously maximized.

\section{Optimization algorithm} 
\label{sec:algorithm}
In this section we describe a particular algorithm we used to look for
the most informative dimensions in order to find the relevant
subspace. We make no claim that our choice of the algorithm is most
efficient. However, it does give reproducible results for different
starting points and spike trains with differences taken to simulate
neural noise. Overall,  choices for an algorithm are broader because the
information $I({\bf v})$ as defined by (\ref{Iv}) is a continuous
function, whose gradient can be computed.  We find (see
Appendix~\ref{app:derivative} for a derivation)
\begin{equation}
\label{grad}
\nabla_{\bf v} I=\int dx P_{\bf v}(x) \left[\langle {\bf s}|x, {\rm spike}\rangle-\langle {\bf s}|x\rangle \right]
\cdot \left[\frac{d}{dx}
\frac{P_{\bf v}(x|{\rm spike})}{P_{\bf v}(x)}\right],
\end{equation}
where 
\begin{equation}
\langle {\bf s}|x,{\rm spike}\rangle={1\over{P(x|{\rm spike})}}\int d {\bf s} \, {\bf s}\delta(x-{\bf s}\cdot
{\bf v})P({\bf
  s}|{\rm spike}),
\end{equation}
and similarly for $\langle {\bf s}|x\rangle$. Since information does
not change with the length of the vector, we have ${\bf v} \cdot \nabla_{\bf v}I =0 $, as
also can be seen  directly from Eq.~(\ref{grad}).

As an optimization algorithm, we have used a combination of gradient
ascent and simulated annealing algorithms: successive line
maximizations were done along the direction of the gradient
\citep{recipes}.  During line maximizations, a point with a smaller
value of information was accepted according to Boltzmann statistics,
with probability $\propto \exp[(I( {\bf v}_{i+1})-I( {\bf v}_i))/T]$.
The effective temperature $T$ is reduced by factor of $1-\epsilon_T$
upon completion of each line maximization.  Parameters of the
simulated annealing algorithm to be adjusted are the starting
temperature $T_0$ and the cooling rate $\epsilon_T$, $\Delta
T=-\epsilon_T T$. When maximizing with respect to one vector we used
values $T_0=1$ and $\epsilon_T=0.05$. When maximizing with respect to
two vectors, we either used the cooling schedule with $\epsilon_T=0.005$ and
repeated it several times (4 times in our case) or allowed the
effective temperature $T$ to increase by a factor of 10 upon
convergence to a local maximum (keeping $T\leq T_0$ always), while
limiting the total number of line maximizations.

\begin{figure}[t]
\begin{center}
\vspace*{-0.05in}
\includegraphics[width=4.25in,height=2.2in]{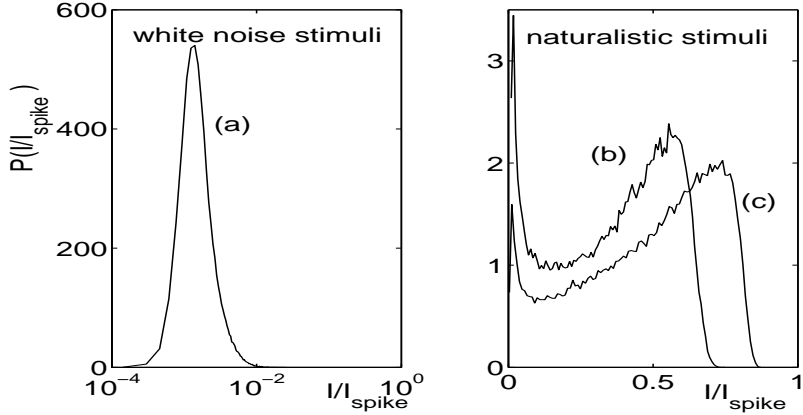}
\end{center}
\vspace*{-0.2in}
\caption{The probability distribution of information values in units
  of the total information per spike in the case of (a) uncorrelated
  binary noise stimuli, (b) correlated Gaussian noise with power
  spectrum of natural scenes, and (c) stimuli derived from natural
  scenes (patches of photos). The distribution was obtained by
  calculating information along $10^5$ random vectors for a model cell
  with one relevant dimension. Note the different scales in the two
  panels.}
\label{fig:random}
\vspace*{-0.1in}
\end{figure}

The problem of maximizing a function often is related to the problem
of making a good initial guess.  It turns out, however, that the
choice of a starting point is much less crucial in cases where the
stimuli are correlated. To illustrate this point we plot in
Fig.~\ref{fig:random} the probability distribution of information along random directions
${\bf v}$ both for white noise and for naturalistic stimuli in a model with one relevant
dimension.  For uncorrelated stimuli, not only is information equal to
zero for a vector that is perpendicular to the relevant subspace, but
in addition  the derivative is equal to zero. Since a randomly chosen vector
has on average a  small projection on the relevant subspace
(compared to its length) $v_r/|{\bf v}|\sim \sqrt{n/d}$, the
corresponding information can be found by expanding in $v_r/|{\bf
  v}|$:
\begin{equation}
\label{inf_values}
I\approx \frac{v_r^2}{2|{\bf v}|^2}\int dx P_{\hat{e}_{ir}}(x) \left(\frac{P'_{\hat{e}_{ir}}(x)}{P_{\hat{e}_{ir}}(x)}\right)^2
[\langle {\bf s}\hat e_r |{\rm spike}\rangle-\langle {\bf s}\hat e_r\rangle]^2
\end{equation}
where  vector $ v=v_r\hat{e}_r+v_{ir}\hat{e}_{ir}$ is decomposed in
its components inside and outside the RS, respectively.
The average information for a random vector is, therefore, $\sim
(\langle v_r^2 \rangle/|{\bf v}|^2) =K/D$.

In cases where stimuli are drawn from a Gaussian ensemble with
correlations, an expression for the information values has a similar
structure to (\ref{inf_values}). To see this, we  transform to Fourier space and
normalize each Fourier component  by the square root of the power spectrum $S({\bf k})$.
In this new basis, both the vectors $\{e_i\}$, $1\leq i \leq K$, forming the RS and
the randomly chosen vector ${\bf v}$ along which information is being
evaluated are to be multiplied by $\sqrt{S({\bf k})}$. Thus, if we now
substitute for the dot product $v_r^2$ the convolution weighted by the
power spectrum, $ \sum_i^K({\bf v} *\hat e_i)^2$, where
\begin{equation}
\label{weighted}
{\bf v} *\hat e_i=\frac{\sum_{\bf k} v({\bf k})\hat
  e_i({\bf k}) S({\bf k})}{\sqrt{\sum_{\bf k} v^2({\bf k}) S({\bf
      k})}\sqrt{\sum_{\bf k} \hat e_i^2({\bf k}) S({\bf k})}},
\end{equation}
then Eq.~(\ref{inf_values}) will describe information values along
randomly chosen vectors ${\bf v}$ for correlated Gaussian stimuli with
the power spectrum $S({\bf k})$. Even though both $v_r$ and $v({\bf
  k})$ are Gaussian variables with variance $\sim 1/D$, the weighted
convolution has not only a much larger variance but is also strongly
non--Gaussian [the non--Gaussian character is  due to the
normalizing factor $\sum_{\bf k} v^2({\bf k}) S({\bf k})$ in the denominator of
Eq.~(\ref{weighted})].  As for the variance, it can be estimated as
$<({\bf v}*\hat e_1)^2> =4\pi/ \ln^2 D$, in cases where stimuli are
taken as patches of correlated Gaussian noise with the two-dimensional
power spectrum $S({\bf k})=A/k^2$. The large values of the weighted
dot product ${\bf v} *\hat e_i$, $1\leq i\leq K$  result not only in significant
information values along a randomly chosen vector, but also in large
magnitudes  of the derivative $\nabla I$, which is no longer dominated by
noise, contrary to the case of uncorrelated stimuli.  In the end, we find that
randomly choosing one of the presented frames as a starting guess is
sufficient.

\section{Results}
\label{sec:results}
\begin{figure}[t]
\begin{center}
\vspace*{-0.2in}
\includegraphics[width=3.25in]{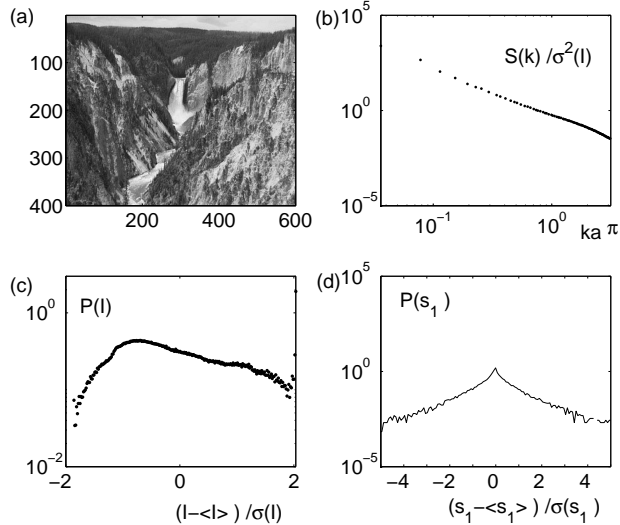}
\end{center}
\vspace*{-0.3in}
\caption{Statistical properties of the visual stimulus ensemble.
Panel (a) shows one of the photos. Stimuli would be 30x30 patches
taken from the overall photograph. In panel (b) we show the power
spectrum, in units of light intensity variance $\sigma^2(I)$,
averaged over orientation as a function of dimensionless wave vector
$ka$, where $a$ is the pixel size.
  (c) The probability distribution of light intensity in units of
$\sigma(I)$. (d) The probability distribution of projections between
stimuli and a Gabor filter, also in units of the corresponding
standard deviation $\sigma(s_1)$.}
\label{fig:statvisual}
\vspace*{-0.12in}
\end{figure}

We tested the scheme of looking for the most informative dimensions on
model neurons that respond to stimuli derived from natural scenes and
sounds. As visual stimuli we used scans across natural scenes, which
were taken as black and white photos digitized to 8 bits with no
corrections made for the camera's light intensity transformation
function.  Some statistical properties of the stimulus set are shown
in Fig.~\ref{fig:statvisual}. Qualitatively, they reproduce the known
results on the statistics of natural scenes
\citep{Ruderman,Ruderman94,Dong95,Simoncelli01}.  Most important
properties for this study are strong spatial correlations, as evident
from the power spectrum S(k) plotted in panel (b), and deviations of
the probability distribution from a Gaussian one. The non--Gaussian
character can be seen in panel (c), where the

\noindent 
probability distribution of intensities is shown, and in panel (d)
which shows the distribution of projections on a Gabor filter [in what
follows the units of projections, such as $s_1$, will be given in
units of the corresponding standard deviations]. Our goal is to
demonstrate that even though the correlations present in the ensemble
are non--Gaussian, they can be removed successfully from the estimate
of vectors defining the RS.

\subsection{A model simple cell}
\begin{figure}[t]
\begin{center}
\vspace*{-0.3in}
\includegraphics[width=3.4in]{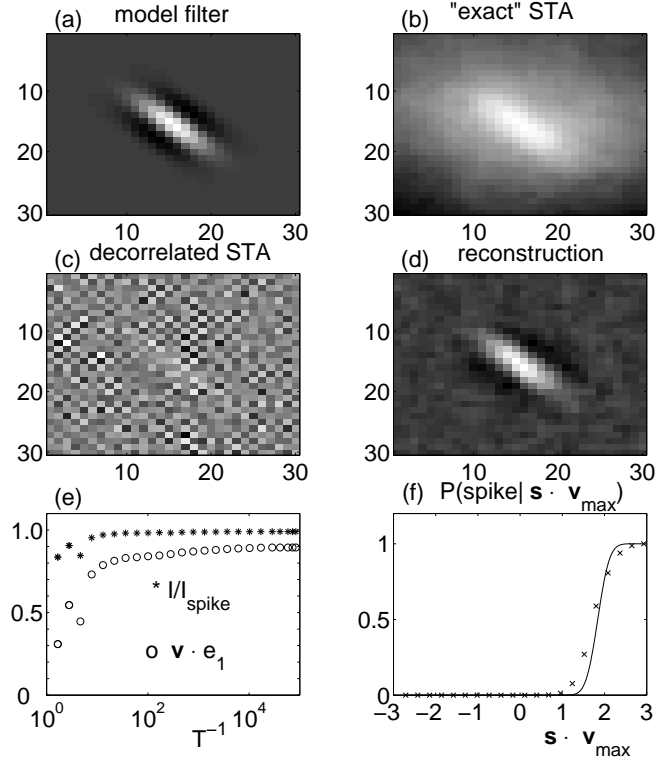}
\end{center}
\vspace*{-0.3in}
\caption{Analysis of a model simple cell with RF shown in (a). The
  ``exact'' spike--triggered average $ {\bf v}_{\rm\tiny sta}$ is shown
  in (b).  Panel (c) shows an attempt to remove correlations according
  to reverse correlation method, $C_{a\, priori}^{-1} {\bf v}_{\rm
  sta}$; (d) the normalized vector $ \hat v_{\rm max}$ found by maximizing
  information; (e) convergence of the algorithm according to
  information $I({\bf v})$ and projection $\hat v \cdot \hat e_1$
  between normalized vectors as a function of inverse effective
  temperature $T^{-1}$. (f) The probability of a spike $P({\rm
  spike}|{\bf s} \cdot \hat v_{max})$ (crosses) is compared to $P({\rm
  spike}|s_1)$ used in generating spikes (solid line).  Parameters of
  the model are $\sigma= 0.31$ and $\theta= 1.84$, both given in units
  of standard deviation of $s_1$, which is also the units for $x$-axis
  in panel (f).}
\label{fig:simple}
\vspace*{-0.1in}
\end{figure}

Our first example is based on the properties of simple cells found in
the primary visual cortex.  A model phase and orientation sensitive
cell has a single relevant dimension $\hat e_1$ shown in
Fig.~\ref{fig:simple}(a).  A given stimulus ${\bf s}$ leads to a spike if
the projection $s_1={\bf s} \cdot \hat e_1$ reaches a threshold value
$\theta$ in the presence of noise:
\begin{equation}
\label{model}
\frac{P({\rm spike}|{\bf s})}{P({\rm spike})}\equiv f(s_1)=\langle H ( s_1-\theta +\xi)\rangle,
\end{equation} 
where a Gaussian random variable $\xi$ of variance $\sigma^2$ models
additive noise, and the function $H(x)=1$ for $x>0$, and zero
otherwise. Together with the RF $\hat e_1$, the parameters
$\theta$ for threshold and the noise variance $\sigma^2$ determine the
input--output function.

When the spike--triggered average (STA), or reverse correlation
function, is computed from the responses to correlated stimuli, the
resulting vector will be broadened due to spatial correlations present
in the stimuli (see Fig.~\ref{fig:simple}b).  For stimuli that are
drawn from a Gaussian probability distribution, the effects of
correlations could be removed by multiplying ${\bf v}_{\rm sta}$ by
the inverse of the {\it a priori} covariance matrix, according to the
reverse correlation method, ${\hat v}_{Gaussian\,est} \propto C_{a\,
  priori}^{-1} {\bf v}_{\rm sta}$, Eq.~(\ref{rcm}). However this
procedure tends to amplify noise.  To separate errors due to neural
noise from those due to the non--Gaussian character of correlations,
note that in a model the effect of neural noise on our estimate of
the STA can be eliminated by averaging the presented stimuli weighted
with the exact firing rate, as opposed to using a histogram of
responses to estimate $P({\rm spike}|{\bf s})$ from a finite set of
trials.  We have used this ``exact'' STA,
\begin{equation}
\label{exact}
{\bf v}_{\rm sta} = \int d{\bf s}\, {\bf s} P({\bf s} | {\rm spike})
= {1\over{P({\rm spike})}}\int d {\bf s} P({\bf s})\, {\bf s} P({\rm spike}|{\bf s}) ,
\end{equation} 
in calculations presented in Figs.~\ref{fig:simple}~(b) and (c).  Even
with this noiseless STA (the equivalent of collecting an infinite data
set), the standard decorrelation procedure is not valid for
non--Gaussian stimuli and nonlinear input--output functions,  as discussed in detail in Appendix~\ref{app:non}.  The
result of such a decorrelation in our example is shown in
Fig.~\ref{fig:simple}(c).  It clearly is missing some of the structure
in the model filter, with projection $\hat e_1\cdot {\hat
v}_{Gaussian\,est} \approx 0.14$. The discrepancy is not due to neural
noise or finite sampling, since the ``exact'' STA was decorrelated;
the absence of noise in the exact STA also means that there would be
no justification for smoothing the results of the decorrelation.  The
discrepancy between the true receptive field and the decorrelated STA
increases with the strength of nonlinearity in the input--output
function.

In contrast, it is possible to obtain a good estimate of the relevant
dimension $\hat e_1$ by maximizing information directly, see panel
(d).  A typical progress of the simulated annealing algorithm with
decreasing temperature $T$ is shown in Fig.~\ref{fig:simple}(e).
There we plot both the information along the vector, and its
projection on $\hat e_1$. We note that while information $I$ remains
almost constant, the value of projection continues to improve.
Qualitatively this is because the probability distributions depend
exponentially on information.  The final value of projection depends
on the size of the data set, as discussed below. In the example shown in
Fig.~\ref{fig:simple} there were $\approx 50,000$ spikes with average
probability of spike $\approx 0.05$ per frame, and the reconstructed
vector has a projection ${\hat v}_{max}\cdot \hat e_1= 0.920\pm 0.006$.
Having estimated the RF, one can proceed to sample the nonlinear
input-output function. This is done by constructing histograms for
$P({\bf s} \cdot {\hat v}_{\rm max})$ and $P({\bf s} \cdot {\hat
  v}_{\rm max}|{\rm spike})$ of projections onto vector $\hat v_{\rm
  max}$ found by maximizing information, and taking their ratio, as in
Eq.~(\ref{bayes}). In Fig.~\ref{fig:simple}(f) we compare $P({\rm
  spike}|{\bf s} \cdot {\hat v}_{\rm max})$ (crosses) with the
probability $P({\rm spike}|s_1)$ used in the model (solid line).

\subsection{Estimated deviation from the optimal dimension}

When information is calculated from a finite data set, the
(normalized) vector ${\hat v}$ which maximizes $I$ will deviate from
the true RF $\hat e_1$. The deviation $\delta {\bf v}={\hat v}-\hat
e_1$ arises because the probability distributions are estimated from
experimental histograms and differ from the distributions found in the
limit of infinite data size. For a simple cell, the quality of
reconstruction can be characterized by the projection ${\hat v} \cdot
\hat e_1=1- \frac{1}{2} \delta {\bf v}^2$, where both ${\hat v}$ and
$\hat e_1$ are normalized, and $\delta {\bf v}$ is by definition
orthogonal to $\hat e_1$.  The deviation $\delta {\bf v} \sim A^{-1}
\nabla I$, where $A$ is the Hessian of information. Its structure is
similar to that of a covariance matrix:
\begin{equation}
A_{ij}=\frac{1}{\ln 2}\int dx P(x|{\rm spike})\left (\frac{d}{dx} \ln \frac{P(x|{\rm spike})}{P(x)}\right)^2( \langle s_is_j|x\rangle-\langle s_i|x\rangle \langle s_j|x\rangle).
\end{equation}

When averaged over possible outcomes of $N$ trials, the gradient of
information is zero for the optimal direction. Here in order to
evaluate $\langle \delta {\bf v}^2\rangle={\rm Tr}[A^{-1}\langle
\nabla I \nabla I^{T}\rangle A^{-1}] $, we need to know the variance
of the gradient of $I$.  Assuming that the probability of generating a
spike is independent for different bins, we can estimate $\langle \nabla
I_i \nabla I_j \rangle\sim A_{ij}/( N_{\rm spike}\ln 2)$. Therefore an
expected error in the reconstruction of the optimal filter is
inversely proportional to the number of spikes. The corresponding
expected value of the projection between the reconstructed vector and the
relevant direction $\hat e_1$ is given by:
\begin{equation}
\label{error}
{\hat v} \cdot \hat e_1 \approx 1-\frac{1}{2}\langle \delta {\bf v}^2\rangle= 1-\frac{{\rm Tr}' [ A^{-1}] }{2 N_{\rm
spike}\ln 2}, 
\end{equation}
where ${\rm Tr}'$ means that the trace is taken in the subspace
orthogonal to the model filter\footnote{ \normalsize By definition $\delta v_1=
\delta {\bf v} \cdot \hat e_1=0$, and therefore $\langle \delta v_1^2
\rangle \propto A_{11}^{-1}$ is to be subtracted from $\langle \delta
{\bf v}^2\rangle \propto {\rm Tr}[A^{-1}]$. Because $\hat e_1$ is an
eigenvector of $A$ with zero eigenvalue, $A_{11}^{-1}$ is infinite.
Therefore the proper treatment is to take the trace in the subspace
orthogonal to $\hat e_1$.}.  The estimate (\ref{error}) can be
calculated without knowledge of the underlying model, it is $\sim
D/(2N_{\rm spike})$.  This behavior should also hold in cases where
the stimulus dimensions are expanded to include time. The errors are
expected to increase in proportion to the increased dimensionality. In
the case of a complex cell with two relevant dimensions, the error is
expected to be twice that for a cell with single relevant dimension,
also discussed in section~\ref{sub:complex}. 

\begin{figure}[t]
\begin{center}
\includegraphics[width=3.2in]{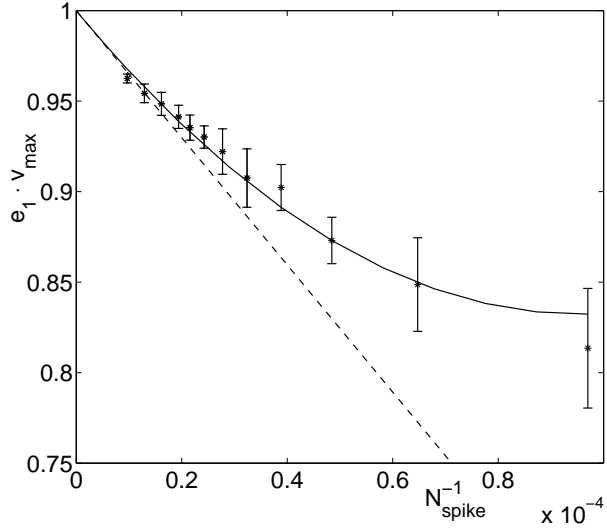}
\end{center}
\vspace*{-0.3in}
\caption{Projection of vector $\hat v_{\rm max}$ that maximizes information
  on RF $\hat e_1$ is plotted as a function of the number of spikes.
  The solid line is a quadratic fit in $1/N_{\rm spike}$, and the
  dashed line is the leading linear term in $1/N_{\rm spike}$. This
  set of simulations was carried out for a model visual neuron with
  one relevant dimension from Fig.~\protect\ref{fig:simple}(a) and the
  input/output function (\protect\ref{model}) with parameter values
  $\sigma\approx 0.61\sigma(s_1)$, $\theta\approx 0.61\sigma(s_1)$. For
  this model neuron, the linear approximation for the expected error
  is applicable for $N_{\rm spike} \protect\agt 30,000$.
}
\label{fig:scaling}
\end{figure}

We emphasize that the error estimate according to Eq.~(\ref{error}) is
of the same order as errors of the reverse correlation method when it
is applied for Gaussian ensembles.  The latter are given by $({\rm
Tr}[C^{-1}]-C_{11}^{-1})/[2N_{\rm spike} \langle f^{'2}(s_1)\rangle]$.
Of course, if the reverse correlation method were to be applied to the
non--Gaussian ensemble, the errors would be larger.  In
Fig.~\ref{fig:scaling} we show the result of simulations for various
numbers of trials, and therefore $N_{\rm spike}$. The average
projection of the normalized reconstructed vector ${\hat v}$ on the RF
$\hat e_1$ behaves initially as $1/N_{\rm spike}$ (dashed line). For
smaller data sets, in this case $N_{\rm spikes} \alt 30,000$,
corrections $\sim N_{\rm spikes}^{-2}$ become important for estimating
the expected errors of the algorithm.  Happily these corrections have
a sign such that smaller data sets are {\em more} effective than one
might have expected from the asymptotic calculation. This can be
verified from the expansion ${\hat v}\cdot \hat e_1= \left[1-\delta
{\bf v}^2\right]^{-1/2} \approx 1- {1\over 2}\langle \delta {\bf
v}^2\rangle + {3\over 8} \langle \delta {\bf v}^4\rangle$, where only
the first two terms where taken into account in Eq.~\ref{error}.

\subsection{A model complex cell}
\label{sub:complex}
\begin{figure}[t]
\begin{center}
\includegraphics[width=3.05in]{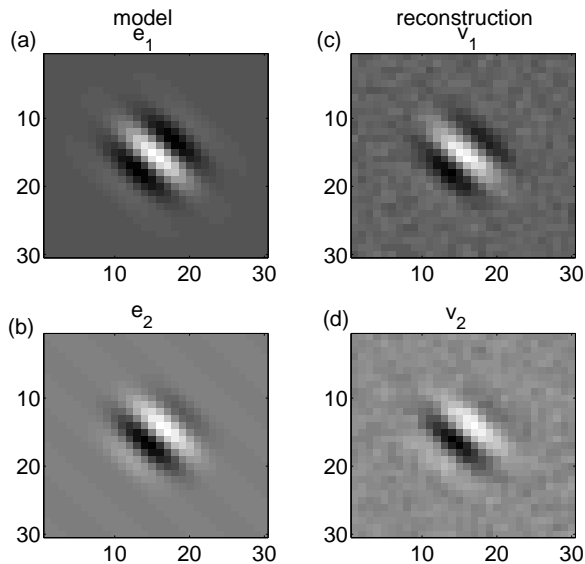}
\end{center}
\vspace*{-0.3in}
\caption{Analysis of a model complex cell with relevant dimensions
$\hat e_1$ and $\hat e_2$ shown in (a) and (b). Spikes are
generated according to an ``OR'' input-output function
$f(s_1,s_2)$ with the threshold 
$\theta\approx 0.61\sigma(s_1)$ and noise standard
deviation $\sigma=0.31\sigma(s_1)$. 
In panels (c) and (d), we show vectors ${\bf
v}_1$ and ${\bf v}_2$ found by maximizing information $I({\bf
v}_1,{\bf v}_2)$.
}
\label{fig:complex}
\end{figure}

A sequence of spikes from a model cell with two relevant dimensions
was simulated by projecting each of the stimuli on vectors that differ
by $\pi/2$ in their spatial phase, taken to mimic properties of
complex cells, as in Fig.~\ref{fig:complex}. A particular frame leads
to a spike according to a logical OR, that is if either $|s_1|$ or
$|s_2|$ exceeds a threshold value $\theta$ in the presence of
noise, where $s_1={\bf s} \cdot \hat e_1$, $s_2={\bf s}
\cdot \hat e_2$. Similarly to (\ref{model}),
\begin{equation}
\frac{P({\rm spike}| {\bf s})}{P({\rm spike})}=f(s_1,s_2)=\langle
H(|s_1|-\theta-\xi_1) \;\vee\;
H(|s_2|-\theta-\xi_2)\rangle\,,
\end{equation}
where $\xi_1$ and $\xi_2$ are independent Gaussian variables.  The
sampling of this input--output function by our particular set of
natural stimuli is shown in Fig.~\ref{fig:complex}(c).
 As is well known, reverse correlation fails in this case because the
spike--triggered average stimulus is zero, although with Gaussian
stimuli the spike--triggered covariance method would recover the
relevant dimensions \citep{Touryan02}.  Here we show that searching for maximally
informative dimensions allows us to recover the relevant subspace even
under more natural stimulus conditions.

We start by maximizing information with respect to one vector.
Contrary to the result Fig.~\ref{fig:simple}(e) for a simple cell, one
optimal dimension recovers only about 60\% of the total information
per spike [Eq.~(\ref{exp_ispike})].  Perhaps surprisingly, because of
the strong correlations in natural scenes, even a projection onto a
random vector in the $D\sim 10^3$ dimensional stimulus space has a
high probability of explaining 60\% of total information per spike, as
can be seen in Fig.~\ref{fig:random}.  We therefore go on to maximize
information with respect to two vectors.  As a result of maximization,
we obtain two vectors ${\bf v}_1$ and ${\bf v}_2$ shown in
Fig.~\ref{fig:complex}. The information along them is $I({\bf v}_1,{\bf
v}_2)\approx 0.90$ which is within the range of information values
obtained along different linear combinations of the two model vectors
$I(\hat e_1,\hat e_2)/I_{\rm spike}=0.89\pm 0.11$. Therefore, the description of
neuron's firing in terms of vectors ${\bf v}_1$ and ${\bf v}_2$ is
complete up to the noise level, and we do not have to look for extra
relevant dimensions. Practically, the number of relevant dimensions
can be determined by comparing $I({\bf v}_1,{\bf v}_2)$ to either
$I_{\rm spike}$ or $I({\bf v}_1,{\bf v}_2,{\bf v}_3)$, the later being
the result of maximization with respect to three vectors
simultaneously. As mentioned in the Introduction, information along
set a of vectors does not increase when extra dimensions are added to
the relevant subspace. Therefore, if $I({\bf v}_1,{\bf v}_2)=I({\bf
v}_1,{\bf v}_2,{\bf v}_3)$ (again, up to the noise level), then this
means that there are only 2 relevant dimensions. Using $I_{\rm spike}$
for comparison with $I({\bf v}_1,{\bf v}_2)$ has the advantage of not
having to look for an extra dimension, which can be computationally
intensive. However $I_{\rm spike}$ might be subject to larger systematic
bias errors than $I({\bf v}_1,{\bf v}_2,{\bf v}_3)$.

Vectors ${\bf v}_1$ and ${\bf v}_2$ obtained by maximizing $I({\bf
v}_1,{\bf v}_2)$ are not exactly orthogonal, and are also rotated with
respect to $\hat e_1$ and $\hat e_2$. However, the quality of
reconstruction, as well as the value of information $I({\bf v}_1,{\bf
v}_2)$, is independent of a particular choice of basis with the
RS. The appropriate measure of similarity between the two planes is
the dot product of their normals.  In the example of
Fig.~\ref{fig:complex}, $\hat n_{(\hat e_1,\hat e_2)}\cdot \hat
n_{({\bf v}_1,{\bf v}_2)} =0.82 \pm
0.07$,
 where $\hat n_{(\hat e_1,\hat
e_2)}$ is a normal to the plane passing through vectors $\hat e_1$ and
$\hat e_2$.  Maximizing information with respect to two dimensions
requires a significantly slower cooling rate, and consequently longer
computational times. However, the expected error in the
reconstruction, $1-\hat n_{(\hat e_1,\hat e_2)}\cdot \hat n_{({\bf
v}_1,{\bf v}_2)}$, scales as $ 1/N_{\rm spike}$ behavior, similarly to
(\ref{error}), and is roughly twice that for a simple cell given the
same number of spikes.
We make vectors ${\bf v}_1$ and ${\bf v}_2$
orthogonal to each others upon completion of the algorithm.

\subsection{A model auditory neuron with one relevant dimension}

\begin{figure}[t]
\vspace*{-0.1in}
\begin{center}
\includegraphics[width=3.65in]{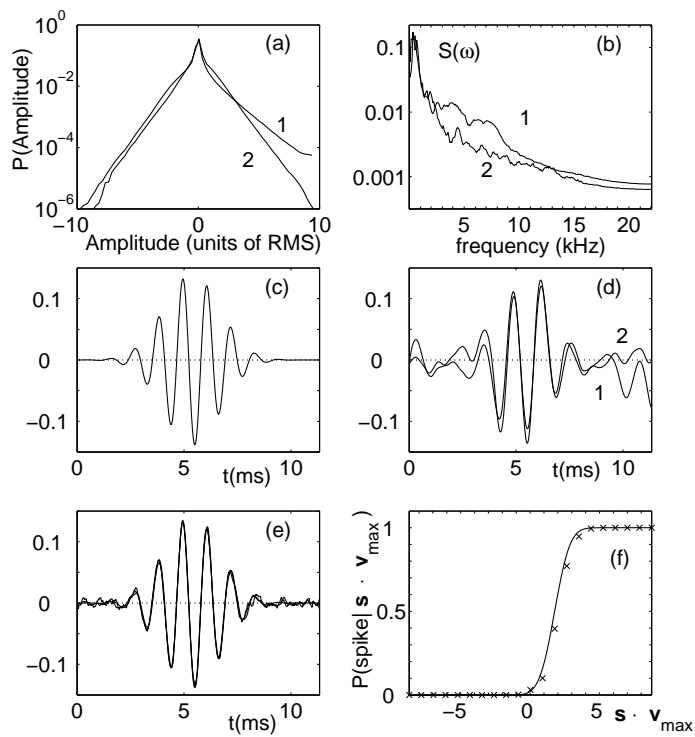}
\end{center}
\vspace*{-0.2in}
\caption{A model auditory neuron is probed by two natural 
  ensembles of stimuli: a piece of English prose (1) and a piece of of
  Russian prose (2) . The size of the stimulus ensemble was the same
  in both cases, and the sampling rate was 44.1 kHz. (a) The
  probability distribution of the sound pressure amplitude in units of
  standard deviation for both ensembles is strongly non--Gaussian. (b)
  The power spectra for the two ensembles. (c) The relevant vector
  of the model neuron, of dimensionality $D=500$. (d) The STA is
  broadened in both cases, but differs among the two cases due to
  differences in the power spectra of the two ensembles. (e) Vectors
  that maximize information for either of the ensembles overlap almost
  perfectly with each other, and with the model filter from (a), which
  is also replotted here from (c). 
(f) The probability of a spike $P({\rm spike}|{\bf s} \cdot \hat
v_{max})$ (crosses) is compared to $P({\rm spike}|s_1)$ used in
generating spikes (solid line). The input-output function had
parameter values $\sigma\approx 0.9\sigma(s_1)$ and $\theta \approx 1.8
\sigma(s_1)$.  }
\label{fig:cochlear}
\vspace*{-.2in}
\end{figure}

Because stimuli ${\bf s}$ are treated as vectors in an abstract space,
the method of looking for the most informative dimensions can be
applied equally well to auditory as well as to visual neurons.  Here
we illustrate the method by considering a model auditory neuron with
one relevant dimension, which is shown in Fig.~\ref{fig:cochlear}(c)
and is taken to mimic the properties of cochlear neurons. The model
neuron is probed by two ensembles of naturalistic stimuli: one is a
recording of a native Russian speaker reading a piece of Russian
prose, and the other one is a recording of a piece of English prose
read by a native English speaker. Both of the ensembles are
non--Gaussian and exhibit amplitude distributions with long, nearly
exponential tails, cf.  Fig.~\ref{fig:cochlear}(a), which are
qualitatively similar to those of light intensities in natural scenes
\citep{Voss75,Ruderman94}.  However, the power spectrum is different
in the two cases, as can be seen in Fig.~\ref{fig:cochlear}(b). The
differences in the correlation structure in particular lead to
different STAs across the two ensembles, cf. panel (d). Both of the
STAs also deviate from the model filter shown in panel (c).

Despite differences in the probability distributions $P({\bf s})$, it
is possible to recover the relevant dimension of the model neuron by
maximizing information.  In panel (e) we show the two most informative
vectors found by running the algorithm for the two ensembles, 
and replot the model filter from (c) to show that the three vectors
overlap almost perfectly. 
Thus different non--Gaussian correlations can be
successfully removed to obtain an estimate of the relevant dimension.
If the most informative vector changes with the stimulus
ensemble, this can be interpreted as caused by adaptation to the
probability distribution.

\section{Summary}
 
Features of the stimulus that are most relevant for generating the
response of a neuron can be found by maximizing information between
the sequence of responses and the projection of stimuli on trial
vectors within the stimulus space. Calculated in this manner,
information becomes a function of direction in stimulus space.  Those
vectors that maximize the information and account for the total
information per response of interest span the relevant subspace.  The
method allows multiple dimensions to be found. The reconstruction of
the relevant subspace is done without assuming a particular form of
the input--output function. It can be strongly nonlinear within the
relevant subspace, and is estimated from experimental histograms for
each trial direction independently.  Most importantly, this method can
be used with any stimulus ensemble, even those that are strongly
non--Gaussian as in the case of natural signals. We have illustrated
the method on model neurons responding to natural scenes and sounds.
We expect the current implementation of the method to be most useful
in cases where several most informative vectors ($\leq 10$, depending
on their dimensionality) are to be analyzed for neurons probed by
natural scenes. This technique could be particularly useful in describing
sensory processing in poorly understood regions of higher level sensory
cortex (such as visual areas V2, V4 and IT and auditory cortex beyond A1)
where white noise stimulation is known to be less effective than
naturalistic stimuli.   

\section*{Acknowledgments}
We thank K.~D.~Miller for many helpful discussions. Work at UCSF was
supported in part by the Sloan and Swartz Foundations and by a
training grant from the NIH. Our collaboration began at the Marine
Biological Laboratory in a course supported by grants from NIMH and
the Howard Hughes Medical Institute.

\appendix

\section{Limitations of the reverse correlation method}
\label{app:non}
Here we examine what sort of deviations one can expect when applying
the reverse correlation method to natural stimuli even in the model
with just one relevant dimension. There are two factors that, when
combined, invalidate the reverse correlation method: the non--Gaussian
character of correlations and the nonlinearity of the input/output
function \citep{Ringach97}. In its original formulation \citep{deBoer}, the neuron is
probed by white noise and the relevant dimension $\hat e_1$ is given
by the STA $\hat e_1 \propto \langle {\bf s} r({\bf s})\rangle$. If
the signals are not white, i.e. the covariance matrix $C_{ij}=\langle
s_is_j\rangle$ is not a unit matrix, then the STA   is a broadened
version of the original filter $\hat e_1$. This can be seen by noting
that for any function $F({\bf s})$ of Gaussian variables $\{s_i\}$ the
identity holds:
\begin{equation}
\label{ward}
 \langle s_i F({\bf s}) \rangle = \langle s_i
s_j\rangle \langle \partial_{s_j} F({\bf s})\rangle, \quad \partial_j\equiv \partial_{s_j}.
\end{equation}
When property (\ref{ward}) is applied to the vector components of the STA, $\langle
s_ir({\bf s})\rangle = C_{ij}\langle\partial_jr({\bf s})\rangle$.
Since we work within the assumption that the firing rate is a
(nonlinear) function of projection onto one filter $\hat
e_1$, $r({\bf s})=r(s_1)$, the later average is proportional to the
model filter itself, $\langle \partial_jr\rangle= {\hat
e_{1j}}\langle r'(s_1)\rangle$. Therefore, we arrive at the
prescription of the reverse correlation method
\begin{equation}
\label{rcm}
\hat e_{1i} \propto [C^{-1}]_{ij}\langle s_j r({\bf s})\rangle .
\end{equation}
The Gaussian property is necessary in order to represent the STA as a
convolution of the covariance matrix $C_{ij}$ of the stimulus ensemble
and the model filter.  To understand how the reconstructed
vector obtained according to Eq.~(\ref{rcm}) deviates from the
relevant one, we consider weakly non--Gaussian stimuli, with the
probability distribution
\begin{equation}
P_{nG}({\bf s})={1\over Z} P_0({\bf s}) e^{\epsilon H_1({\bf s})} ,
\end{equation}
where $P_0({\bf s})$ is the
Gaussian probability distribution with covariance matrix $C$, and the
normalization factor $Z=\langle e^{\epsilon H_1({\bf s})}\rangle$.
The function $H_1$ describes deviations of the probability distribution
from Gaussian, and therefore we will set $\langle s_i H_1 \rangle=0$
and $\langle s_i s_j H_1 \rangle=0$, since these averages can be
accounted for in the Gaussian ensemble. In what follows we will keep
only the first order terms in perturbation parameter $\epsilon$. Using
the property (\ref{ward}), we find the STA to be given by

\begin{equation}
\label{1st}
\langle s_i r\rangle_{nG}=\langle s_i s_j \rangle \left[\langle \partial_j r\rangle +\epsilon
 \langle r\partial_j (H_1)\rangle\right],
\end{equation}
where averages are taken with respect to the Gaussian distribution.
Similarly, the covariance matrix $C_{ij}$ evaluated with respect to
the non--Gaussian ensemble is given by:
\begin{equation}
C_{ij}={1\over Z} \langle s_i s_j e^{\epsilon H_1}\rangle =\langle s_i s_j\rangle +\epsilon \langle s_i s_k\rangle \langle s_j\partial_k(H_1)\rangle 
\end{equation}
so that to the first order in $\epsilon$, $\langle s_i
s_j\rangle=C_{ij}-\epsilon C_{ik} \langle s_j \partial_k
(H_1)\rangle$. Combining this with Eq.~(\ref{1st}), we get
\begin{equation}
\label{expan2}
\langle s_i r\rangle_{nG}={\rm const} \times C_{ij}\hat e_{1j} +\epsilon C_{ij}\langle \left(r-s_1\langle r'\rangle\right)\partial_j(H_1)\rangle.
\end{equation}
The second term in (\ref{expan2}) prevents the application of the
reverse correlation method for non--Gaussian signals. Indeed, if we multiply
the STA (\ref{expan2}) with the inverse of the {\it a priori}
covariance matrix $C_{ij}$ according to the reverse correlation method
(\ref{rcm}), we no longer obtain the RF $\hat e_1$.  The deviation of
the obtained answer from the true RF increases with $\epsilon$, which
measures the deviation of the probability distribution from Gaussian.
Since natural stimuli are known to be strongly non--Gaussian, this
makes the use of the reverse correlation problematic when analyzing
neural responses to natural stimuli.

\begin{figure}[t]
\begin{center}
\vspace*{-0.6in}
\includegraphics[width=4.25in]{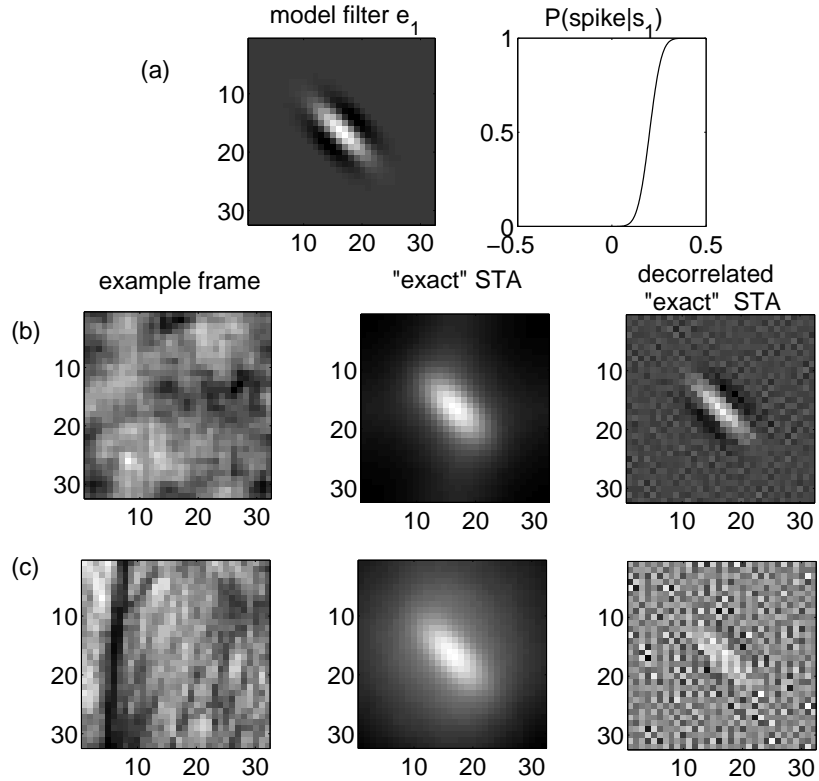}
\end{center}
\vspace*{-0.1in}
\caption{The non--Gaussian character of correlations present in natural scenes invalidates 
  the reverse correlation method for neurons with a nonlinear
  input-output function. Here, a model visual neuron has one relevant
  dimension $\hat e_1$ and the nonlinear input/output function, shown
  in (a).  The ``exact'' STA is used (\protect\ref{exact}) to separate
  effects of neural noise from alterations introduced by the method.
  The decorrelated ``exact'' STA is obtained by multiplying the
  ``exact'' STA by the inverse of the covariance matrix, according to
  Eq.~(\protect\ref{rcm}).  (b) Stimuli are taken from a correlated
  Gaussian noise ensemble.  The effect of correlations in STA can be
  removed according to Eq.~(\protect\ref{rcm}).  When patches of photos
  are taken as stimuli (c) for the same model neuron as in (b), the
  decorrelation procedure gives an altered version of the model
  filter. The two stimulus ensembles have the same covariance matrix.
  }
\label{fig:gauss_exact.eps}
\vspace*{-0.1in}
\end{figure}

\begin{figure}[t]
\begin{center}
\vspace*{-0.8in}
\includegraphics[width=3.9in, height=4.1in]{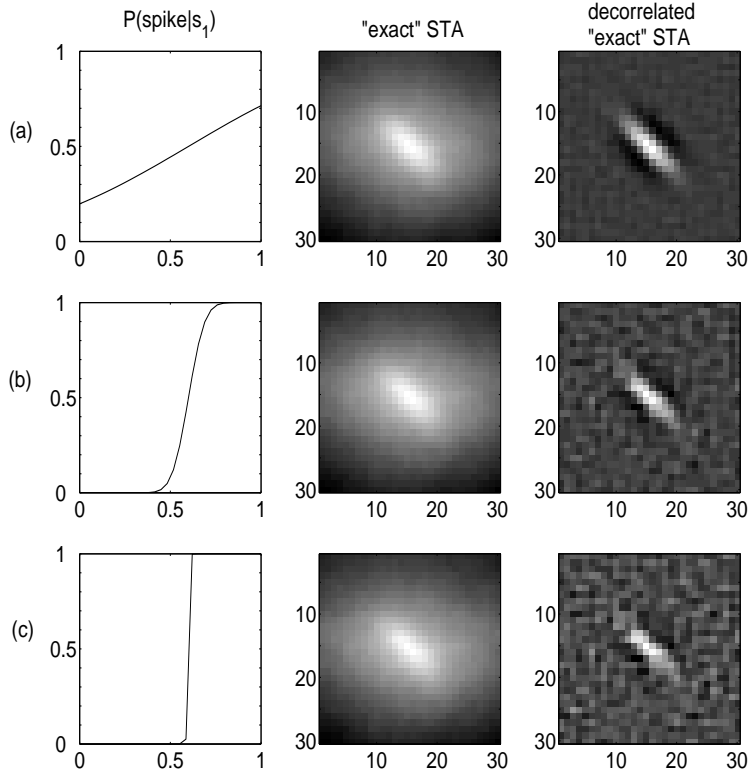}
\end{center}
\vspace*{-0.2in}
\caption{Application of the reverse correlation method to a model
  visual neuron with one relevant dimension $\hat e_1$ and a threshold
  input/output function of decreasing values of noise variance
  $\sigma/\sigma(s_1)s\approx6.1,\,0.61,\,0.06$ in (a), (b), and
  (c) respectively. The model $P({\rm spike}|s_1)$ becomes effectively
  linear when signal-to noise ratio is small. The reverse correlation
  can be used together with natural stimuli, if the input-output
  function is linear.  Otherwise, the deviations between the
  decorrelated STA and the model filter increase with nonlinearity of
  $P({\rm spike}|s_1)$.}
\label{fig:gauss_exact2.eps}
\vspace*{0.02in}
\end{figure}

The difference in applying the reverse correlation to stimuli drawn
from a correlated Gaussian ensemble vs. a non--Gaussian one is
illustrated in Figs.~\ref{fig:gauss_exact.eps}~(b) and (c). In the
first case, shown in (b), stimuli are drawn from a correlated Gaussian
ensemble with the covariance matrix equal to that of natural images.
In the second case, shown in (c), the patches of photos are taken as
stimuli.  The STA is broadened in both cases. Even though the two-point correlations
are just as strong in the case of Gaussian stimuli as they are in the
natural stimuli ensemble, Gaussian correlations can be successfully
removed from the STA according to Eq.~(\ref{rcm}) to obtain the model
filter.  On the contrary, an attempt to use reverse correlation with
natural stimuli results in an altered version of the model filter. We
reiterate that for this example the apparent noise in the decorrelated vector is not
due to neural noise or finite datasets, since the ``exact'' STA has
been used (\ref{exact}) in all calculations presented in
Figs.~\ref{fig:gauss_exact.eps} and \ref{fig:gauss_exact2.eps}.

The reverse correlation method gives the correct answer for any distribution
of signals if the probability of generating a spike is a linear
function of $s_i$, since then the second term in Eq. (\ref{expan2}) is
zero. In particular, a linear input-output relation could arise due to
a neural noise whose variance is much larger than the variance of the
signal itself.  This point is illustrated in
Figs.~\ref{fig:gauss_exact2.eps}~(a), (b), and (c), where the reverse
correlation method is applied to a threshold input--output function at low, moderate, and high signal-to-noise ratios.  For small
signal-to-noise ratios where the noise standard deviation is similar
to that of projections $s_1$, the threshold nonlinearity in the
input-output function is masked by noise, and is effectively linear.
In this limit, the reverse correlation can be applied with the exact
STA.  However, for experimentally calculated STA at low
signal-to-noise ratios the decorrelation procedure results in strong
noise amplification. On the other hand, at higher signal-to-noise
ratios decorrelation fails due to the nonlinearity of the input-output
function in accordance with (\ref{expan2}).

\section{Maxima of  $I(v)$:  what do they mean?}
\label{app:max}
The relevant subspace of dimensionality $K$ can be found by maximizing
information simultaneously with respect to $K$ vectors. The result of
maximization with respect to a number of vectors that is less than the
true dimensionality of the relevant subspace may produce vectors which
have components in the irrelevant subspace. This happens only in the
presence of correlations in stimuli. As an illustration, we consider
the situation where the dimensionality of the relevant subspace $K=2$,
and vector $\hat e_1$ describes the most informative direction {\it
  within} the relative subspace. We show here that even though the
gradient of information is perpendicular to both $\hat e_1$ and $\hat
e_2$, it may have components outside the relevant subspace. Therefore the
vector $v_{\rm max}$ that corresponds to the maximum of $I( v)$ will
then lie outside the relevant subspace.  We recall from Eq. (10) that
\begin{equation}
\nabla I(\hat e_1)= \int d s_1 P(s_1) \frac{d}{d s_1} \frac
{P(s_1|{\rm spike})}{P(s_1)} (\langle{\bf s}|s_1, {\rm spike}\rangle -\langle {\bf s}|s_1\rangle), 
\end{equation}
We can rewrite the conditional averages $\langle {\bf s}|s_1\rangle
= \int ds_2 P(s_1,s_2)\langle {\bf
  s}|s_1,s_2\rangle/P(s_1)$ and \\
$\langle {\bf s}|s_1, {\rm spike}\rangle = \int ds_2
f(s_1,s_2)P(s_1,s_2)\langle {\bf s}|s_1,s_2\rangle/P(s_1|{\rm
  spike})$, so that
\begin{eqnarray}
\label{grad2}
\hspace*{-0.4in}\nabla I(\hat e_1)= \int d s_1 ds_2 P(s_1,s_2)\langle {\bf
s}|s_1,s_2\rangle  \frac {P({\rm
spike}|s_1,s_2) -P({\rm
spike}|s_1)}{P({\rm spike})} \frac{d}{ds_1}  \ln \frac {P(s_1|{\rm
spike})}{P(s_1)} .
\end{eqnarray}
Because we assume that the vector $\hat e_1$ is the most informative
within the relevant subspace, $\hat e_1\nabla I=\hat e_2\nabla I=0$,
so that the integral in (\ref{grad2}) is zero for those directions in
which the component of the vector $\langle {\bf s}|s_1,s_2\rangle$
changes linearly with $s_1$ and $s_2$. For uncorrelated stimuli this
is true for all directions, so that the most informative vector within
the relevant subspace is also the most informative in the overall
stimulus space. In the presence of correlations, the gradient may have
non-zero components along some irrelevant directions if projection of
the vector $\langle {\bf s}|s_1,s_2\rangle$ on those directions is not
a linear function of $s_1$ and $s_2$.  By looking for a maximum of
information we will therefore be driven outside the relevant
subspace. The deviation of $v_{\rm max}$ from the relevant subspace is
also proportional to the strength of the dependence on the second
parameter $s_2$, because of the factor $[P(s_1,s_2|{\rm
spike})/P(s_1,s_2) -P(s_1|{\rm spike})/P(s_1)]$ in the integrand.

\section{The gradient of information}
\label{app:derivative}
According to the expression (\ref{Iv}), the information $I({\bf v})$
depends on the vector ${\bf v}$ only through the probability
distributions $P_{\bf v}(x)$ and $P_{\bf v}(x|{\rm spike})$. Therefore we can express the gradient of information in terms of gradients of those probability distributions:
\begin{eqnarray}
\label{gradi}
\nabla_{\bf v}I  = \frac{1}{\ln 2} \int dx \left[\ln\frac{P_{\bf v}(x|{\rm spike})}{P_{\bf v}(x)}
\nabla_{\bf v}(P_{\bf v}(x|{\rm spike}))  -\frac{P_{\bf v}(x|{\rm spike})}{P_{\bf v}(x)}\nabla_{\bf v}(P_{\bf v}(x))\right],
\end{eqnarray} 
where we took into account that $\int dx P_{\bf v}(x|{\rm spike})=1$
and does not change with ${\bf v}$.  To find gradients of the probability distributions, we note that
\begin{eqnarray}
\label{gradPx}
\hspace*{-0.2in} \nabla_{\bf v}P_{\bf v}(x)=\nabla_{\bf v}\left[\int d{\bf s} P({\bf s}) \delta(x-{\bf s}\cdot {\bf v})\right]=-\int d{\bf s} P({\bf s}) {\bf s}\delta'(x-{\bf s}\cdot {\bf v}) = -\frac{d}{dx}\left[ p(x)\langle {\bf s} |x \rangle\right], 
\end{eqnarray}
and analogously  for $P_{\bf v}(x|{\rm spike})$:
\begin{equation}
\label{gradPxt} 
\nabla_{\bf v}P_{\bf v}(x|{\rm spike})=  -\frac{d}{dx}\left[ p(x|{\rm
     spike})\langle {\bf s} |x, {\rm spike} \rangle\right].
\end{equation}
Substituting expressions (\ref{gradPx}) and (\ref{gradPxt}) into Eq.~(\ref{gradi}) and integrating once by parts we obtain:
\[
\nabla_{\bf v} I=\int dx P_{\bf v}(x) \left[\langle {\bf s}|x, {\rm spike}\rangle-\langle {\bf s}|x\rangle \right]
\cdot \left[\frac{d}{dx}
\frac{P_{\bf v}(x|{\rm spike})}{P_{\bf v}(x)}\right],
\]
which is the expression (\ref{grad}) of the main text.

\begin{thebibliography}{46}
\expandafter\ifx\csname natexlab\endcsname\relax\def\natexlab#1{#1}\fi
\expandafter\ifx\csname url\endcsname\relax
  \def\url#1{{\tt #1}}\fi
\expandafter\ifx\csname urlprefix\endcsname\relax\def\urlprefix{URL }\fi

\bibitem[{Ag\"{u}era y Arcas et~al.(2003)}]{Arcas03}
Ag\"{u}era y Arcas, B., Fairhall, A.~L. \& Bialek, W. (2003).
\newblock Computation in a single neuron: Hodgkin and Huxley revisited.
\newblock {\em Neural Comp.\/}, {\em 15\/}, 1715--1749.



\bibitem[{Baddeley et~al.(1997)Baddeley, Abbott, Booth, Sengpiel, Freeman,
  Wakeman, \& Rolls}]{Baddeley97}
Baddeley, R., Abbott, L.~F., Booth, M. C.~A., Sengpiel, F., Freeman, T.,
  Wakeman, E.~A., \& Rolls, E.~T. (1997).
\newblock Responses of neurons in primary and inferior temporal visual cortices
  to natural scenes.
\newblock {\em Proc. R. Soc. Lond. B\/}, {\em 264\/}, 1775--1783.

\bibitem[{Barlow(1961)}]{Barlow61}
Barlow, H. (1961).
\newblock Possible principles underlying the transformations of sensory images.
\newblock In {\em Sensory Communication\/}, edited by W.~Rosenblith. MIT Press,
  Cambridge,  217--234.

\bibitem[{Barlow(2001)}]{Barlow01}
Barlow, H. (2001).
\newblock Redundancy reduction revisited.
\newblock {\em Network: Comput. Neural Syst.\/}, {\em 12\/}, 241--253.

\bibitem[{Bialek(2002)}]{leshouches_notes}
Bialek, W. (2002).
\newblock Thinking about the brain.
\newblock In {\em Physics of Biomolecules and Cells\/}, edited by H.~Flyvbjerg,
  F.~J{\"{u}}licher, P.~Ormos, \& F.~David. EDP Sciences, Les Ulis;
  Springer-Verlag, Berlin,  485--577.
\newblock See also physics/0205030.\footnote{\normalsize Where available we give references to the physics e--print archive, which may be found at http://arxiv.org/abs/*/*; thus Bialek (2002) is available at
http://arxiv.org/abs/physics/0205030.  Published papers may differ from the versions posted to the archive.}


\bibitem[{Bialek \& de~Ruyter~van Steveninck(2003)}]{rob+bill-feature}
Bialek, W., \& de~Ruyter~van Steveninck, R.~R. (2003).
\newblock Features and dimensions: Motion estimation in fly vision.
\newblock In preparation.

\bibitem[{Brenner et~al.(2000{\natexlab{a}})Brenner, Bialek, \& de~Ruyter~van
  Steveninck}]{brenner00-adapt}
Brenner, N., Bialek, W., \& de~Ruyter~van Steveninck, R.~R.
  (2000{\natexlab{a}}).
\newblock Adaptive rescaling maximizes information transmission.
\newblock {\em Neuron\/}, {\em 26\/}, 695--702.

\bibitem[{Brenner et~al.(2000{\natexlab{b}})Brenner, Strong, Koberle, Bialek,
  \& de~Ruyter~van Steveninck}]{B00a}
Brenner, N., Strong, S.~P., Koberle, R., Bialek, W., \& de~Ruyter~van
  Steveninck, R.~R. (2000{\natexlab{b}}).
\newblock Synergy in a neural code.
\newblock {\em Neural Computation\/}, {\em 12\/}, 1531--1552.
\newblock See also physics/9902067.

\bibitem[{Chichilnisky(2001)}]{Chichilnisky01}
Chichilnisky, E.~J. (2001).
\newblock A simple white noise analysis of neuronal light responses.
\newblock {\em Network: Comput. Neural Syst\/}, {\em 12\/}, 199--213.

\bibitem[{Creutzfeldt \& Northdurft(1978)}]{Creutzfeldt78}
Creutzfeldt, O.~D., \& Northdurft H.~C. (1978).
\newblock Representation of complex visual stimuli in the brain.
\newblock {\em Naturwissenshaften\/}, {\em 65\/}, 307--318.


\bibitem[{Cover \& Thomas(1991)}]{Cover_Thomas}
Cover, T.~M., \& Thomas, J.~A. (1991).
\newblock {\em Information theory\/}.
\newblock John Wiley \& Sons, INC., New York.

\bibitem[{de~Boer \& Kuyper(1968)}]{deBoer}
de~Boer, E., \& Kuyper, P. (1968).
\newblock Triggered correlation.
\newblock {\em IEEE Trans. Biomed. Eng.\/}, {\em 15\/}, 169--179.

\bibitem[{Dimitrov \& Miller(2001)}]{JPMiller01}
Dimitrov, A.~G., \& Miller, J.~P. (2001).
\newblock Neural coding and decoding: communication channels and quantization.
\newblock {\em Network: Comput. Neural Syst.\/}, {\em 12\/}, 441--472.

\bibitem[{Dong \& Atick(1995)}]{Dong95}
Dong, D.~W., \& Atick, J.~J. (1995).
\newblock Statistics of natural time-varying images.
\newblock {\em Network: Comput. Neural Syst.\/}, {\em 6\/}, 345--358.

\bibitem[{Fairhall et~al.(2001)Fairhall, Lewen, Bialek, \& de~Ruyter~van
  Steveninck}]{Fairhall01}
Fairhall, A.~L., Lewen, G.~D., Bialek, W., \& de~Ruyter~van Steveninck, R.~R.
  (2001).
\newblock Efficiency and ambiguity in an adaptive neural code.
\newblock {\em Nature\/}, 787--792.

\bibitem[{Kara et~al.(2000)Kara, Reinagel, \& Reid}]{Kara00}
Kara, P., Reinagel, P., \& Reid, R.~C. (2000).
\newblock Low response variability in simultaneously recorded retinal,
  thalamic, and cortical neurons.
\newblock {\em Neuron\/}, {\em 27\/}, 635--646.

\bibitem[{Lewen et~al.(2001)Lewen, Bialek, \& de~Ruyter~van
  Steveninck}]{Lewen01}
Lewen, G.~D., Bialek, W., \& de~Ruyter~van Steveninck, R.~R. (2001).
\newblock Neural coding of naturalistic motion stimuli.
\newblock {\em Network: Comput. Neural Syst.\/}, {\em 12\/}, 317--329.
\newblock See also physics/0103088.

\bibitem[{Mainen \& Sejnowski(1995)}]{MainenSejnowski}
Mainen, Z.~F., \& Sejnowski, T.~J. (1995).
\newblock Reliability of spike timing in neocortical neurons.
\newblock {\em Science\/}, {\em 268\/}, 1503--1506.

\bibitem[{Paninski(2003{\natexlab{a}})}]{Paninski_NIPS03}
Paninski, L. (2003{\natexlab{a}}).
\newblock Convergence properties of three spike--triggered analysis techniques.
\newblock {\em Network: Compt. in Neural Systems\/}, {\em 14}, 437--464.

\bibitem[{Paninski(2003{\natexlab{b}})}]{Paninski_NC03}
Paninski, L. (2003{\natexlab{b}}).
\newblock Estimation of entropy and mutual information.
\newblock {\em Neural Computation\/}, {\em 15\/}, 1191-1253.

\bibitem[{Panzeri \& Treves(1996)}]{Panzeri96}
Panzeri, S., \& Treves, A. (1996).
\newblock Analytical estimates of limited sampling biases in different
  information measures.
\newblock {\em Network: Comput. Neural Syst.\/}, {\em 7\/}, 87--107.

\bibitem[{Pola et~al.(2002)Pola, Schultz, Petersen, \& Panzeri}]{Pola02}
Pola, G., Schultz, S.~R., Petersen, R., \& Panzeri, S. (2002).
\newblock A practical guide to information analysis of spike trains.
\newblock In {\em Neuroscience databases: a practical guide\/}, edited by
  R.~Kotter. Kluwer Academic Publishers,  137--152.

\bibitem[{Press et~al.(1992)Press, Teukolsky, Vetterling, \&
  Flannery}]{recipes}
Press, W.~H., Teukolsky, S.~A., Vetterling, W.~T., \& Flannery, B.~P. (1992).
\newblock {\em Numerical Recipes in C: The Art of Scientific Computing\/}.
\newblock Cambridge University Press, Cambridge.

\bibitem[{Reinagel \& Reid(2000)}]{Reinagel00}
Reinagel, P., \& Reid, R.~C. (2000).
\newblock Temporal coding of visual information in the thalamus.
\newblock {\em J. Neurosci.\/}, {\em 20\/}, 5392--5400.

\bibitem[{Rieke et~al.(1995)Rieke, Bodnar, \& Bialek}]{Rieke95}
Rieke, F., Bodnar, D.~A., \& Bialek, W. (1995).
\newblock Naturalistic stimuli increase the rate and efficiency of information
  transmission by primary auditory afferents.
\newblock {\em Proc. R. Soc. Lond. B\/}, {\em 262\/}, 259--265.

\bibitem[{Rieke et~al.(1997)Rieke, Warland, de~Ruyter~van Steveninck, \&
  Bialek}]{Rieke_book}
Rieke, F., Warland, D., de~Ruyter~van Steveninck, R.~R., \& Bialek, W. (1997).
\newblock {\em Spikes: Exploring the neural code\/}.
\newblock MIT Press, Cambridge.

\bibitem[{Ringach et~al.(1997)Ringach, Sapiro, \& Shapley}]{Ringach97}
Ringach, D.~L., Sapiro, G., \& Shapley, R. (1997).
\newblock A subspace reverse-correlation technique for the study of visual neurons.
\newblock {\em Vision Res.\/}, {\em 37\/}, 2455--2464.

\bibitem[{Ringach et~al.(2002)Ringach, Hawken, \& Shapley}]{Ringach02}
Ringach, D.~L., Hawken, M.~J., \& Shapley, R. (2002).
\newblock Receptive field structure of neurons in monkey visual cortex revealed
  by stimulation with natural image sequences.
\newblock {\em Journal of Vision\/}, {\em 2\/}, 12--24.

\bibitem[{Rolls et~al.(2003)Rolls, Aggelopoulos, \& Zheng}]{Rolls03}
Rolls, E.~T., Aggelopoulos, N.~C., \& Zheng, F. (2003).
\newblock The receptive fields of inferior temporal cortex neurons in natural
  scenes.
\newblock {\em J. Neurosci.\/}, {\em 23\/}, 339--348.

\bibitem[{Ruderman(1994)}]{Ruderman94}
Ruderman, D.~L. (1994).
\newblock The statistics of natural images.
\newblock {\em Network: Compt. Neural Syst.\/}, {\em 5\/}, 517--548.

\bibitem[{Ruderman \& Bialek(1994)}]{Ruderman}
Ruderman, D.~L., \& Bialek, W. (1994).
\newblock Statistics of natural images: scaling in the woods.
\newblock {\em Phys. Rev. Lett.\/}, {\em 73\/}, 814--817.


\bibitem[{de~Ruyter~van Steveninck \& Bialek(1988)}]{stcov}
de~Ruyter~van Steveninck, R.~R., \& Bialek, W. (1988).
\newblock Real-time performance of a movement-sensitive neuron in the blowfly
  visual system: coding and information transfer in short spike sequences.
\newblock {\em Proc. R. Soc. Lond. B\/}, {\em 265\/}, 259--265.

\bibitem[{de~Ruyter~van Steveninck et~al.(2001)de~Ruyter~van Steveninck, Borst,
  \& Bialek}]{canberra}
de~Ruyter~van Steveninck, R.~R., Borst, A., \& Bialek, W. (2001).
\newblock Real time encoding of motion: Answerable questions and questionable
  answers from the fly's visual system, 279-306.
\newblock In {\em Motion Vision: 
  Computational, Neural and Ecological Constraints\/}, edited by J.~M. Zanker,
  \& J.~Zeil. Springer--Verlag, Berlin.
\newblock See also physics/0004060.

\bibitem[{de~Ruyter~van Steveninck et~al.(1997)de~Ruyter~van Steveninck, Lewen,
  Strong, Koberle, \& Bialek}]{deRuyter97}
de~Ruyter~van Steveninck, R.~R., Lewen, G.~D., Strong, S.~P., Koberle, R., \&
  Bialek, W. (1997).
\newblock Reproducibility and variability in neural spike trains.
\newblock {\em Science\/}, {\em 275\/}, 1805--1808.
\newblock See also cond-mat/9603127.

\bibitem[{Schwartz et~al.(2002)Schwartz, Chichilnisky, \&
  Simoncelli}]{Schwartz02}
Schwartz, O., Chichilnisky, E.~J., \& Simoncelli, E. (2002).
\newblock Characterizing neural gain control using spike--triggered covariance.
\newblock In {\em Advances in Neural Information Processing\/}, edited by T.~G.
  Dietterich, S.~Becker, \& Z.~Ghahramani, vol.~14, 279-306.

\bibitem[{Sen et~al.(2001)Sen, Theunissen, \& Doupe}]{Sen01}
Sen, K., Theunissen, F.~E., \& Doupe, A.~J. (2001).
\newblock Feature analysis of natural sounds in the songbird auditory
  forebrain.
\newblock {\em J. Neurophysiol.\/}, {\em 86\/}, 1445--1458.

\bibitem[{Simoncelli \& Olshausen(2001)}]{Simoncelli01}
Simoncelli, E., \& Olshausen, B.~A. (2001).
\newblock Natural image statistics and neural representation.
\newblock {\em Annu. Rev. Neurosci.\/}, {\em 24\/}, 1193--1216.

\bibitem[{Smirnakis et~al.(1996)Smirnakis, Berry, Warland, Bialek, \&
  Meister}]{Smirnakis96}
Smirnakis, S.~M., Berry, M.~J., Warland, D.~K., Bialek, W., \& Meister, M.
  (1996).
\newblock Adaptation of retinal processing to image contrast and spatial scale.
\newblock {\em Nature\/}, {\em 386\/}, 69--73.

\bibitem[{Smyth et~al.(2003)Smyth, Willmore, Baker, Thompson, \& Tolhurst}]{Smyth03}
Smyth, D., Willmore, B., Baker, G.~E., Thompson, I.~D., \& Tolhurst, D.~J. (2003).
\newblock The receptive-field organization of simple cells in primary visual cortex of ferrets under natural scene stimulation.
\newblock {\em J. Neurosci.\/}, {\em 23\/}, 4746--4759.


\bibitem[{Stanley et~al.(1999)Stanley, Li, \& Dan}]{Stanley99}
Stanley, G.~B., Li, F.~F., \& Dan, Y. (1999).
\newblock Reconstruction of natural scenes from ensemble responses in the
  lateral geniculate nucleus.
\newblock {\em J. Neurosci.\/}, {\em 19\/}, 8036--8042.

\bibitem[{Strong et~al.(1998)Strong, Koberle, de~Ruyter~van Steveninck, \&
  Bialek}]{entropy}
Strong, S.~P., Koberle, R., de~Ruyter~van Steveninck, R.~R., \& Bialek, W.
  (1998).
\newblock Entropy and information in neural spike trains.
\newblock {\em Phys. Rev. Lett.\/}, {\em 80\/}, 197--200.

\bibitem[{Theunissen et~al.(2000)Theunissen, Sen, \& Doupe}]{Theunissen00}
Theunissen, F.~E., Sen, K., \& Doupe, A.~J. (2000).
\newblock Spectral-temporal receptive fields of nonlinear auditory neurons
  obtained using natural sounds.
\newblock {\em J. Neurosci.\/}, {\em 20\/}, 2315--2331.

\bibitem[{Tishby et~al.(1999)Tishby, Pereira, \& Bialek}]{Tishby}
Tishby, N., Pereira, F.~C., \& Bialek, W. (1999).
\newblock The information bottleneck method.
\newblock In {\em Proceedings of the 37th Allerton Conference on Communication,
  Control and Computing\/}, edited by B.~Hajek, \& R.~S. Sreenivas. University
  of Illinois, 368-377.
\newblock See also physics/0004057.

\bibitem[{Touryan et~al.(2002)}]{Touryan02}
 Touryan J., Lau, B., \& Dan, Y. (2002).
\newblock Isolation of relevant visual features from
random stimuli for cortical complex cells.  \newblock {\em J. Neurosci.\/}, {\em  22\/}, 10811--10818.

\bibitem[{Treves \& Panzeri(1995)}]{Treves}
Treves, A., \& Panzeri, S. (1995).
\newblock The upward bias in measures of information derived from limited data
  samples.
\newblock {\em Neural Comp.\/}, {\em 7\/}, 399--407.


\bibitem[{von~der Twer \& Macleod(2001)}]{Twer01}
von~der Twer, T., \& Macleod, D. I. A. (2001).
\newblock Optimal nonlinear codes for the perception of natural colours.
\newblock {\em Network: Comput. Neural Syst.\/}, {\em 12\/}, 395--407.


\bibitem[{Vickers et~al.(2001)Vickers, Christensen, Baker, \&
  Hildebrand}]{Vickers01}
Vickers, N.~J., Christensen, T.~A., Baker, T., \& Hildebrand, J.~G. (2001).
\newblock Odour-plume dynamics influence the brain's olfactory code.
\newblock {\em Nature\/}, {\em 410\/}, 466--470.

\bibitem[{Vinje \& Gallant(2000)}]{Vinje00}
Vinje, W.~E., \& Gallant, J.~L. (2000).
\newblock Sparse coding and decorrelation in primary visual cortex during
  natural vision.
\newblock {\em Science\/}, {\em 287\/}, 1273--1276.

\bibitem[{Vinje \& Gallant(2002)}]{Vinje02}
Vinje, W.~E., \& Gallant, J.~L. (2002).
\newblock Natural stimulation of the nonclassical receptive field increases
  information transmission efficiency in V1.
\newblock {\em J. Neurosci.\/}, {\em 22\/}, 2904--2915.

\bibitem[{Voss \& Clarke(1975)}]{Voss75}
Voss, R.~F., \& Clarke, J. (1975).
\newblock '1/f noise' in music and speech.
\newblock {\em Nature\/}, 317--318.

\bibitem[{Weliky et~al.(2003)Weliky, Fiser, Hunt, \& Wagner}]{Weliky03}
Weliky, M., Fiser, J., Hunt, R., \& Wagner, D.~N. (2003).
\newblock Coding of natural scenes in primary visual cortex.
\newblock {\em Neuron\/}, {\em 37\/}, 703--718.

\end{thebibliography}

\end{document}